\documentclass[a4paper,11pt,reqno]{amsart}

\usepackage{amsmath,amsfonts,amssymb,amscd,enumerate}

\def\a{\alpha}
\def\p{\partial}
\def\A{\mathcal{A}}
\def\R{\mathbb{R}}
\def\C{\mathbb{C}}

\def\g{\mathfrak{g}}
\def\tp{\otimes}
\def\Mk{\mathcal{M}_\kappa}
\def\Dk{\mathcal{D}_\kappa}

\begin{document}

\title{Differential structure on $\kappa$-Minkowski space, and $\kappa$-Poincar\'{e} algebra}

\author{Stjepan Meljanac}
\address[S. Meljanac]{Rudjer Bo\v{s}kovi\'{c} Institute, Bijeni\v{c}ka cesta b.b., 10000 Zagreb, Croatia}
\email{meljanac@irb.hr}
\author{Sa\v{s}a Kre\v{s}i\'{c}-Juri\'{c}}
\address[S. Kre\v{s}i\'{c}-Juri\'{c}]{Faculty of Natural and Mathematical Sciences, University of Split, Teslina 12, 21000 Split, Croatia}
\email{skresic@pmfst.hr}

\thanks{We would like to thank Prof. A. Borowiec for helpful discussions.
This work is supported by the Croatian Ministry of Science, Education and Sports grants no. 098-0000000-2865 and 177-0372794-2816.}
\date{}
\maketitle

\begin{abstract}
We construct realizations of the generators of the $\kappa$-Minkowski space and $\kappa$-Poincar\'{e} algebra as formal power series in the
$h$-adic extension of the Weyl algebra. The Hopf algebra structure of the $\kappa$-Poincar\'{e} algebra related to different realizations is given.
We construct realizations of the exterior derivative and one-forms, and define a differential calculus on $\kappa$-Minkowski space
which is compatible with the action of the Lorentz algebra. In contrast to the conventional bicovariant calculus,
the space of one-forms has the same dimension as the $\kappa$-Minkowski space.
\end{abstract}
\vskip 0.5cm


\section{Introduction}

Recent years have witnessed a growing interest in applications of noncommutative (NC) geometry to a possible unification of quantum field theory
and gravity \cite{Phys-NC-World}--\cite{Chaichian-2}. Current progress in high-energy physics relies in great part on ideas which embody a
modification in the description of spacetime as a continuous geometrical structure. This modification is a natural consequence of the appearance
of a new fundamental length scale known as Planck length \cite{Doplicher}, \cite{Doplicher-2}. The Planck length plays a fundamental role in
loop quantum gravity where a quantization process leads to the area and volume operators having discrete spectra. The minimal values of the
corresponding eigenvalues are proportional to the square and cube of the Planck length, respectively \cite{Rovelli}, \cite{Thiemann}. As a new
fundamental, observer-independent quantity, the Planck length is incorporated into kinematic theory within the framework of the doubly special
relativity (DSR) theory \cite{Amelino-Camelia-1}, \cite{Amelino-Camelia-2}. In DSR there exist two observer-independent scales, velocity
(identified with the speed of light) and length or mass (expected to be the Planck length or Planck mass). The Minkowski spacetime is deformed
into a noncommutative space for which a mathematical model is provided by the $\kappa$-Minkowski space
\cite{Amelino-Camelia-1}--\cite{Kowalski-Glikman-2}. The symmetry algebra for doubly special relativity is obtained by deforming the ordinary
Poincar\'{e} algebra into a Hopf algebra known as $\kappa$-Poincar\'{e} algebra \cite{Kowalski-Glikman-1}--\cite{Kowalski-Glikman-3}. Different
representations (bases) \cite{Kowalski-Glikman-1} of the $\kappa$-Poincar\'{e} algebra correspond to different versions of the DSR theory.
However, the resulting space-time algebra is independent of the representation \cite{Kowalski-Glikman-1}, \cite{Kowalski-Glikman-3}. Recently,
the $\kappa$-Minkowski NC space in bicrossproduct basis was shown to emerge from considerations of a NC differential structure on a
pseudo-Rimannian manifold \cite{Majid-3}, \cite{Majid-4}.

As a part of a general effort to understand the structure of NC spaces, in this paper we shall be interested in developing
differential calculus on the $\kappa$-Minkowski space. The $\kappa$-Minkowski space was studied by different groups, from both the mathematical
and physical points of view. The construction of differential calculus on the $\kappa$-Minkowski space
was considered by Sitarz in Ref. \cite{Sitarz}. He has shown that there is no fourdimensional bicovariant
differential calculus which is Lorentz covariant. If one requires that both conditions are met, then the space of one-forms becomes
five dimensional. His work was subsequently generalized to $n$ dimensions by Gonera et. al. \cite{Gonera}.
A drawback of this approach is that if one-forms are to be generated
by an action of the exterior derivative on the NC coordinates, then there should be exactly $n$ forms obtained in this way.
There have been several attempts to circumvent this problem in the Euclidean and Minkowski space \cite{Dimitrijevic}--\cite{Bu-Kim}.
In Ref. \cite{Meljanac-1} the authors have constructed a noncommutative version of one-forms on the $\kappa$-Euclidean
space as deformations of ordinary one-forms. The NC forms are obtained by an action of a deformed exterior derivative on NC
coordinates. In Ref. \cite{Bu-Kim}, Bu et. al. constructed a differential algebra on the
$\kappa$-Minkowski space from Jordanian twist of the Weyl algebra and showed that the algebra is closed in four dimensions. In their approach
they extended the $\kappa$-Poincar\'{e} algebra with a dilatation operator and used a coproduct of the Lorentz generators which is different from
the one used in Ref. \cite{Sitarz}.

The present paper is a continuation of previous work on differential forms discussed in Ref. \cite{Meljanac-1}. Here we construct a differential
algebra on the $\kappa$-Minkowski space in which the number of NC one-forms $\xi_\mu$ is equal to the number of NC coordinates $\hat x_\mu$.
This is a key difference between our approach and the one presented in Ref. \cite{Sitarz}. The differential algebra is compatible with an
action of the Lorentz generators $M_{\mu\nu}$ and has the property that the commutator $[\xi_\mu,\hat x_\nu]$ is closed in the space
spanned by the one-forms $\xi_\mu$ alone. The closedness of the commutator is important
since in this case any $k$-form can be written as a linear combination of forms of the type
$f_p(\hat x)\xi_{\mu_1} \xi_{\mu_2}\ldots \xi_{\mu_p}$, $0\leq p\leq k$, where $f_p(\hat x)$ is a monomial in $\hat x_\mu$.

The paper is organized as follows. In section 2 we introduce the aglebra generated by the $\kappa$-Minkowski coordinates
$\hat x_\mu$ and Lorentz generators $M_{\mu\nu}$. We extend this algebra by the momentum operators $p_\mu$ such that
$M_{\mu\nu}$ and $p_\mu$ generate the $\kappa$-deformed Poincar\'{e} algebra. We illustrate by examples that this
extension is not unique, and in a special case it leads to the undeformed Poincar\'{e} algebra. We then study realizations
of the coordinates $\hat x_\mu$ and generators $M_{\mu\nu}$ as formal power series in the
$h$-adic extension of the Weyl algebra. We find a large class of such realizations requiring that the commutator $[M_{\mu\nu},\hat
x_\lambda]$ is of Lie type. These realizations generalize the results from Refs. \cite{Meljanac-2} and \cite{Borowiec-2}. Of particular
importance is the noncovariant realization used in the construction of the differential algebra in section 4. In section 3 we give a
brief description of the Hopf algebra structure of the $\kappa$-Poincar\'{e} algebra based on the realizations found in
section 2. Section 4 deals with differential algebra on the $\kappa$-Minkowski space. We find realizations of the exterior derivative
$\hat d$ whose action on NC coordinates $\hat x_\mu$ leads to deformed one-forms $\xi_\mu$. The one-forms $\xi_\mu$ are constructed
as elements of the $h$-adic extension of a super Weyl algebra. The algebra found using these realizations has
the important property that the commutator $[\xi_\mu,\hat x_\nu]$ is closed in the vector space spanned by one-forms $\xi_\mu$ alone.
Since $\hat x_\mu$, $M_{\mu\nu}$ and $\xi_\mu$ belong to an associative algebra, all graded Jacobi identities are automatically satisfied.
The Jacobi identities allow us to define an action of $M_{\mu\nu}$ on the algebra generated by $\hat x_\mu$ and $\xi_\mu$ which is
compatible with he structure of this algebra. This action is different from the one found by Sitarz in Ref. \cite{Sitarz} and it does not
require introduction of an additional one-form. However, when restricted to the coordinates of the $\kappa$-Minkowski space
it agrees with the action found in Ref. \cite{Sitarz}. We note that in our approach the exterior derivative is not Lorentz-invariant
and one-forms do not transform vector-like under the action of the Lorentz generators.

\section{$\kappa$-Minkowski space with Lorentz and $\kappa$-Poincar\'{e} algebra}

In this section we consider the $\kappa$-Minkowski space with $\kappa$-Poincar\'{e} algebra, and their realizations as formal power
series in the $h$-adic extension of the Weyl algebra. This construction was introduced in Refs. \cite{Meljanac-2} and \cite{Meljanac-3}
for the $\kappa$-deformed Euclidean space.

The $\kappa$-Minkowski space is an algebra generated by NC coordinates $\hat x_0,\hat x_1,\ldots ,\hat x_{n-1}$ satisfying the
commutation relations
\begin{equation}\label{01}
[\hat x_\mu, \hat x_\nu] = i(a_\mu \hat x_\nu-a_\nu \hat x_\mu ), \quad a \in \R^n.
\end{equation}
The coordinates $\hat x_\mu$ generate a Lie algebra with structure constants
$C^\lambda_{\mu\nu}=a_\mu \delta_{\nu\lambda}- a_\nu\delta_{\mu\lambda}$ describing a deformation of the ordinary Minkowski space.
One may view $\hat x_\mu$ as deformations of ordinary commutative coordinates $x_\mu$ in the sense that $\hat x_\mu\to x_\mu$
as $a\to 0$. Let $\mathcal{L}$ denote the Lorentz algebra generated by $M_{\mu\nu}$,
\begin{equation}\label{02}
[M_{\mu\nu},M_{\lambda\rho}] = \eta_{\nu\lambda} M_{\mu\rho}-\eta_{\mu\lambda} M_{\nu\rho}-\eta_{\nu\rho} M_{\mu\lambda} +\eta_{\mu\rho}
M_{\nu\lambda},
\end{equation}
where $\eta=diag(-1,1,\ldots,1)$ is the Minkowski metric. The $\kappa$-Minkowski space $\Mk$ and Lorentz algebra $\mathcal{L}$
can be embedded into a Lie algebra $\g_\kappa$ which contains $\Mk$ and $\mathcal{L}$ as Lie subalgebras, and
$\g_\kappa=\Mk\oplus \mathcal{L}$ as vector spaces. The correct form of the mixed commutator for $M_{\mu\nu}$ and $\hat x_\lambda$ in the
Euclidean case was found in Ref. \cite{Meljanac-2}. For the Minkowski metric it is given by
\begin{equation}\label{03}
[M_{\mu\nu},\hat x_\lambda] = \eta_{\nu\lambda} \hat x_\mu - \eta_{\mu\lambda} \hat x_\nu -ia_\mu M_{\nu\lambda}+ia_\nu M_{\mu\lambda}.
\end{equation}
Since all Jacobi identitites for $M_{\mu\nu}$ and $\hat x_\lambda$ hold, Eqs. \eqref{01}--\eqref{03}
define a Lie algebra strucure on $\g_\kappa$. The algebra $\g_\kappa$ can be extended further by momentum generators $p_\mu$
satisfying the commutation relations
\begin{align}
[p_\mu,p_\nu] &= 0,  \label{94} \\
[p_\mu,\hat x_\nu] &= -iH_{\mu\nu}(p), \label{95} \\
[M_{\mu\nu},p_\lambda] &= G_{\mu\nu\lambda}(p),  \label{96}
\end{align}
where $H_{\mu\nu}$ and $G_{\mu\nu\lambda}$ are real-analytic functions of $p$ which generally depend on the deformation parameter $a$.
We require that $H_{\mu\nu}$ and $G_{\mu\nu\lambda}$ satisfy the classical limit conditions
\begin{equation}
\lim_{a\to 0} H_{\mu\nu}=\eta_{\mu\nu}, \quad \lim_{a\to 0} G_{\mu\nu\lambda}=\eta_{\nu\lambda} p_\mu -\eta_{\mu\lambda} p_\nu.
\end{equation}
Relations \eqref{01}, \eqref{94} and \eqref{95} define a deformed Heisenberg algebra $\mathcal{H}_\kappa$, while relations
\eqref{02}, \eqref{94} and \eqref{96} define a $\kappa$-deformed Poincar\'{e} algebra $\mathcal{P}_\kappa$.

The choice of deformations $H_{\mu\nu}$ and $G_{\mu\nu\lambda}$ must be compatible with
the requirement that $\hat x_\mu$, $M_{\mu\nu}$ and $p_\mu$ satisfy the Jacobi relations. The Jacobi identities for
$\hat x_\mu$ and $p_\mu$ imply that $H_{\mu\nu}$ satisfy a system of partial differential equations (PDE's)
\begin{equation}
\sum_{\a=0}^{n-1} \left(\frac{\p H_{\lambda\nu}}{\p p_\a} H_{\a\mu}-\frac{\p H_{\lambda\mu}}{\p p_\a} H_{\a \nu}\right)
=a_\mu H_{\lambda\nu}-a_\nu H_{\lambda\mu}.
\end{equation}
Similarly, the Jacobi identities for $M_{\mu\nu}$ and $p_\mu$ hold if and only if
\begin{equation}
i\sum_{\a=0}^{n-1} \left(G_{\mu\nu\a}\frac{\p G_{\lambda\rho\sigma}}{\p p_\a}-G_{\lambda\rho\a}\frac{\p G_{\sigma\mu\nu}}{\p p_\a}\right)
=\eta_{\nu\lambda}G_{\mu\rho\sigma}-\eta_{\mu\lambda}G_{\nu\rho\sigma}-\eta_{\nu\rho} G_{\mu\lambda\sigma}+
\eta_{\mu\rho}G_{\nu\lambda\sigma}.
\end{equation}
Furthermore, the remaining Jacobi identity for $\hat x_\mu$, $p_\nu$ and $M_{\lambda\rho}$ can be used to derive a system of PDE's relating
the functions $H_{\mu\nu}$ and $G_{\mu\nu\lambda}$.

It is important to note that the extension of $\g_\kappa$ by the momentum generators $p_\mu$ is not unique since the differential equations
for $H_{\mu\nu}$ and $G_{\mu\nu\lambda}$ admit an infinite family of solutions. For example, one solution
is given by
\begin{align}
H_{\mu\nu}(P) &= \eta_{\mu\nu}\left(aP+\sqrt{1+a^2 P^2}\right)-a_\mu P_\nu, \label{97}  \\
G_{\mu\nu\lambda}(P) &= \eta_{\nu\lambda} P_\mu - \eta_{\mu\lambda} P_\nu  \label{98}
\end{align}
where the scalar product in \eqref{97} is taken with respect the Minkowski metric ($aP=-a_0P_0+\sum_{i=1}^{n-1}a_iP_i$).
By straightforward computation one can check that all Jacobi relations for $\hat x_\mu$, $M_{\mu\nu}$ and $P_\mu$
are satisfied. In view of Eq. \eqref{98} this particular solution yields the undeformed Poincar\'{e} algebra.
Another solution with $a=(a_0,0,\ldots,0)$ is given by
\begin{alignat}{2}
H_{00}(p) &= -\psi(A), \quad &  H_{0j}(p) &= 0,  \label{99} \\
H_{i0}(p) &= -a_0 p_i\, \gamma(A), \quad & H_{ij}(p) &= \delta_{ij}\, \varphi(A),  \label{100}
\end{alignat}
and
\begin{align}
G_{i00}(p) &= -\frac{\psi(A)}{\varphi(A)} p_i, \label{103} \\
G_{i0j}(p) &= \delta_{ij}\varphi(A)
\Big(\frac{1-e^{\Psi(A)}}{a_0}-\frac{a_0}{2}\, \square\, e^{\Psi(A)}\Big)-a_0\frac{\gamma(A)}{\varphi(A)}p_i p_j, \label{104} \\
G_{ij0}(p) &= 0, \quad G_{ijk}(p) = \delta_{jk}p_i-\delta_{ik}p_j, \label{105}
\end{align}
where $A=a_0 p_0$, $\psi$ and $\varphi$ are arbitrary real-analytic functions such that $\psi(0)=\varphi(0)=1$, $\varphi^\prime (0)$
is finite and $\gamma=\frac{\varphi^\prime}{\varphi}\psi+1$. The function $\Psi(A)$ is defined by Eq. \eqref{34} and the
deformed Laplace operator $\square$ is given by Eq. \eqref{58} (with $p_\mu=-i\p_\mu$.) As we shall see shortly, these deformations
are related to realizations disucussed in section 2.1.

The deformed Heisenberg algebra $\mathcal{H}_\kappa$ acts on the subalgebra $\mathcal{M}_\kappa\subset \mathcal{H}_\kappa$ as follows.
Let $1$ denote the unit in $\mathcal{M}_\kappa$ and define the action $\rhd \colon \mathcal{H}_\kappa \times \Mk\to \Mk$ by
\begin{enumerate}[(i)]
\item $\hat x_\mu \rhd f(\hat x) = \hat x_\mu f(\hat x)$,
\item $p_\mu \rhd f(\hat x) = [p_\mu,f(\hat x)]\rhd 1, \quad p_\mu \rhd 1 = 0$,
\item $(ab)\rhd 1 = a\rhd (b\rhd 1)$ for all $a,b\in \mathcal{H}_\kappa$,
\end{enumerate}
for any monomial $f(\hat x)\in \Mk$. The rules (i) and (iii) imply that $f(\hat x)\rhd g(\hat x) = f(\hat x)g(\hat x)$
for all monomials $f(\hat x),g(\hat x) \in \mathcal{M}_\kappa$. Consider now the action of the momentum generator $p_\mu$. Since
$\lim_{a\to 0} H_{\mu\nu}(p)=\eta_{\mu\nu}$ we have $H_{\mu\nu}(p)=\eta_{\mu\nu}+o(p)$. Hence, in view of Eq. \eqref{95} the
action of $p_\mu$ on monomials of order one yields
\begin{equation}
p_\mu \rhd \hat x_\nu = [p_\mu,\hat x_\nu]\rhd 1 = -i\big(\eta_{\mu\nu}+o(p)\big) \rhd 1 = -i\eta_{\mu\nu}.
\end{equation}
Similarly, for monomials of order two we have
\begin{align}
p_\mu \rhd (\hat x_\nu \hat x_\lambda) &= [p_\mu,\hat x_\nu \hat x_\lambda]\rhd 1 = \big([p_\mu,\hat x_\nu]\hat x_\lambda+
\hat x_\nu [p_\mu,\hat x_\lambda]\big)\rhd 1   \notag \\
&= \Big(-i\big(\eta_{\mu\nu}+o(p)\big)\hat x_\lambda - i\hat x_\nu \big(\eta_{\mu\lambda}+o(p)\big)\Big)\rhd 1  \notag \\
&= -i\big(\eta_{\mu\nu}\hat x_\lambda + \eta_{\mu\lambda}\hat x_\nu\big) - i \big(o(p)\hat x_\lambda\big)\rhd 1.
\end{align}
This leads to a deformed Leibniz rule for the action of $p_\mu$. We recognize $-i\big(o(p)\hat x_\lambda\big)\rhd 1$ as a deformation of
the standard Leibniz rule $-i(\eta_{\mu\nu}\hat x_\lambda+\eta_{\mu\lambda}\hat x_\nu)$.
The deformation obviously depends on the function $H_{\mu\nu}(p)$. To illustrate the point consider the deformed Heisenberg algebra
\eqref{99}-\eqref{100} with $\varphi-\psi=1+ia_0\p_0$. Then one finds
\begin{equation}
p_\mu \rhd (\hat x_\nu \hat x_\lambda) = -i(\eta_{\mu\nu} \hat x_\lambda + \eta_{\mu\lambda} \hat x_\nu) +
a_0 \eta_{\mu\nu} \eta_{0\lambda}.
\end{equation}
Therefore, the coproduct $\Delta p_\mu$ induced by the Leibniz rule is also deformed.
In the classical limit as $a\to 0$ the Heisenberg algebra $\mathcal{H}_\kappa$ becomes undeformed and $p_\mu$ obeys the standard Leibniz rule
$p_\mu \rhd (\hat x_\nu \hat x_\lambda) = -i(\eta_{\mu\nu} \hat x_\lambda + \eta_{\mu\lambda}\hat x_\nu)$.
Hence, $\lim_{a\to 0}\Delta p_\mu = \Delta_0 p_\mu$ where $\Delta_0 p_\mu=1\otimes p_\mu + p_\mu \otimes 1$ is the primitive coproduct.
Deformations of the Leibniz rule and coproduct described above are discussed in Refs. \cite{Meljanac-2}, \cite{Meljanac-3} and \cite{Meljanac-4}.

\subsection{Realizations}

In this section we shall study deformations of the Heisenberg and Poincar\'{e} algebras using realizations of the generators
as formal power series in the $h$-adic extension of the Weyl algebra.
We want to represent coordinates $\hat x_\mu$ as deformations of commutative coordinates
$x_\mu$ depending on the parameter $a\in \R^n$ in Eq. \eqref{01}. Let $\mathcal{A}_n$ be the Weyl algebra over the field
of complex numbers $\C$ generated by $x_\mu$ and the differential operators $\p_\mu \equiv \frac{\p}{\p x_\mu}$, $\mu = 0,1,\ldots, n-1$.
The generators of $\A_n$ satisfy the commutation relations
\begin{equation}\label{89}
[x_\mu,x_\nu]=[\p_\mu,\p_\nu]=0, \quad [\p_\mu,x_\nu]=\eta_{\mu\nu}.
\end{equation}
Let $\A_n[[a]]$ denote the $h$-adic extension of $\A_n$. The elements of $\A_n[[a]]$ are formal power series in $a_0,a_1,\ldots ,a_{n-1}$
with coefficients in $\A_n$. Consider a representation of $\hat x_\mu$ as an element of $\A_n[[a]]$ given by
\begin{equation}\label{04}
\hat x_\mu = \sum_{\a=0}^{n-1} x^\a \, \phi_{\a\mu}(\p),
\end{equation}
where $x^\a = \sum_\beta x_\beta\, \eta_{\beta\a}$, and $\phi_{\a\mu}$ is a formal power series in $a_\mu$ with coefficients in the
ring of differential operators $\p_\mu$. We require that $\hat x_\mu \to x_\mu$ as $a\to 0$ which
implies that $\lim_{a\to 0} \phi_{\mu\nu}=\eta_{\mu\nu}$.
A representation \eqref{04} is called a $\phi$-realization of the NC coordinates $\hat x_\mu$.
This realization is compatible with commutation relations \eqref{01} if and only if $\phi_{\mu\nu}$ satisfy the system of PDE's
\begin{equation}\label{93}
\sum_{\beta=0}^{n-1}\left(\frac{\p \phi_{\a\mu}}{\p \p_\beta}\phi_{\beta\nu}-
\frac{\p \phi_{\a\nu}}{\p \p_\beta} \phi_{\beta\mu}\right) = ia_\mu \phi_{\a\nu}-ia_{\nu}\phi_{\a\mu}.
\end{equation}
Given the complexity of Eqs. \eqref{93} the
system is often symplified by assuming that $\phi_{\mu\nu}$ are functions of the commuting variables $A=ia\p$ and $B=a^2\p^2$
where the scalar product is taken with respect to the Minkowski metric ($uv = -u_0 v_0+\sum_{i=1}^{n-1}u_i v_i$).
A large class of such realizations in the Euclidean case was found in Refs. \cite{Meljanac-2}, \cite{Meljanac-3} and \cite{Meljanac-4}.

Consider realizations of the Lorentz generators $M_{\mu\nu}$ and momenta $p_\mu$ given by
\begin{equation}\label{101}
M_{\mu\nu}=\sum_{\a=0}^{n-1} x^{\a} \Gamma_{\mu\nu\a}(\p) \quad \text{and}\quad p_\mu=-i\p_\mu
\end{equation}
where $\Gamma_{\mu\nu\a}$ is a formal power series in $a_\mu$ with coefficients in the ring
of differential operators $\p_\mu$. In the classical limit we require that
$\lim_{a\to 0}\Gamma_{\mu\nu\a}=\eta_{\a\mu}\p_\nu - \eta_{\a\nu}\p_\mu$. The functions $\Gamma_{\mu\nu\a}$
are uniquely determined by the commutation relations \eqref{02}--\eqref{03} and the realization \eqref{04}.
Substituting the realizations for $\hat x_\mu$, $M_{\mu\nu}$ and $p_\mu$ into Eqs. \eqref{95} and \eqref{96}
we find
\begin{equation}\label{102}
\phi_{\mu\nu}(\p)=H_{\mu\nu}(-i\p), \quad \Gamma_{\mu\nu\lambda}(\p)=-iG_{\mu\nu\lambda}(-i\p).
\end{equation}
Thus, if the realization of the momentum generator is fixed by $p_\mu=-i\p_\mu$ there is a one-to-one correspondence
between the realizations of the generators $\hat x_\mu$ and $M_{\mu\nu}$ and deformations of the algebras $\mathcal{H}_\kappa$ and $\Mk$.
In the rest of the paper we assume that the momenta have the fixed realization $p_\mu=-i\p_\mu$.

A key tool in the construction of differential forms to be discussed in section 4 is the shift operator. The shfit
operator $Z$ is an element of $\A_n[[a]]$ defined by the commutation relations
\begin{equation}\label{69}
[Z,\hat x_\mu] = ia_\mu Z, \quad [Z,\p_\mu]=0.
\end{equation}
The first relation in Eq. \eqref{69} implies that conjugation by $Z^n$ shifts the coordinate $\hat x_\mu$ by the amount $ina_\mu$,
\begin{equation}\label{92}
Z^n \hat x_\mu Z^{-n} = \hat x_\mu + ina_\mu, \quad n\in \mathbb{Z}.
\end{equation}
The shift operator also satisfies the relation $\hat x_\mu Z\hat x_\nu = \hat x_\nu Z \hat x_\mu$.

\subsubsection{Natural realization}

Different realizations are obtained by choosing different admissible functions $\phi_{\mu\nu}$. Alternatively, starting from a
fixed realization $\phi_{\mu\nu}$ one can introduce a change of generators of the Weyl algebra, $x_\mu \mapsto X_\mu(x,\p)$ and
$\p_\mu\mapsto D_\mu(\p)$, to obtain new realizations. A class of such transformations called similarity transformations
was described in Ref. \cite{Meljanac-4}.

In this paper we shall consider two types of realizations, noncovariant \cite{Meljanac-2} and a special type of covariant
realizations known as the natural realization \cite{Meljanac-3}. The variables used to express these two types of
realizations in Eq. \eqref{04} will be denoted by $(x_\mu,\p_\mu)$ and $(X_\mu,D_\mu)$, respectively. The natural realization
is given by
\begin{equation}\label{27}
\hat x_\mu = X_\mu Z^{-1}+i(aX)D_\mu,
\end{equation}
where $Z^{-1}$ is the inverse shift operator
\begin{equation}\label{35}
Z^{-1} = -iaD+\sqrt{1-a^2 D^2}.
\end{equation}
One can show that if the NC coordinates are given by Eq. \eqref{27}, then the Lorentz generators have the standard representation
\begin{equation}\label{33}
M_{\mu\nu}=X_\mu D_\nu - X_\nu D_\mu.
\end{equation}
Thus, Eqs. \eqref{27}--\eqref{33} provide the natural realization of the aglebra \eqref{01}--\eqref{03}. Since the realization
of the momentum generators is given by $P_\mu = -iD_\mu$, $M_{\mu\nu}$ and $P_\mu$ generate the undeformed Poincar\'{e}
algebra. Note that the natural realization corresponds to our first example of algebra deformation \eqref{97}--\eqref{98}.
This example is rather special since in a generic realization the Poincar\'{e} algebra is deformed.

\subsubsection{Noncovariant realizations}

In the rest of the paper we shall restrict out attention to deformations of the Minkowski space when $a=(a_0,0,\ldots,0)$.
Then the commutation relations \eqref{01} and \eqref{03} yield
\begin{alignat}{2}
[\hat x_i,\hat x_j] &=0, & \qquad [\hat x_0,\hat x_j] &=ia_0\hat x_j, \label{28} \\
[M_{i0},\hat x_0] &=-\hat x_i+ia_0 M_{i0}, & \qquad [M_{i0},\hat x_k] &= -\delta_{ik}\hat x_0+ia_0 M_{ik}, \label{43} \\
[M_{ij},\hat x_0] &=0, & \qquad [M_{ij},\hat x_k] &= \delta_{jk} \hat x_i - \delta_{ik} \hat x_j.  \label{44}
\end{alignat}
By convention the greek indices run through the set $\{0,1,\ldots ,n-1\}$, and the latin indices run through the subset
$\{1,2,\ldots ,n-1\}$.

A family of noncovariant realizations of $\hat x_\mu$ satisfying the algebra \eqref{28} is given by
\begin{align}
\hat x_0 &= x_0 \psi(A)+ia_0 \Big(\sum_{k=1}^{n-1} x_k \p_k\Big) \gamma(A), \quad \hat x_i = x_i \varphi (A), \label{29} \\
\gamma &= \frac{\varphi^\prime}{\varphi}\psi+1, \label{30}
\end{align}
where $A=-ia_0\p_0$. This family is parametrized by two real-analytic functions $\varphi$ and $\psi$ sastisfying
the initial conditions $\varphi(0)=\psi(0)=1$ and $\varphi^\prime (0)$ is finite.
The shift operator in the noncovariant realization is found to be
\begin{equation}\label{34}
Z=e^{\Psi(A)}, \quad \Psi(A)=\int_0^A \frac{dt}{\psi(t)}.
\end{equation}
For a given realization \eqref{29} we want to find a realization of the Lorentz generators such that
$M_{\mu\nu}$ generate the undeformed Lorentz algebra \eqref{02} and $[M_{\mu\nu},\hat x_\lambda]$ is given by
\eqref{43}--\eqref{44}. The realization of $M_{\mu\nu}$ can be found from the natural realization \eqref{33} using
the transformation of variables $(x_\mu,\p_\mu)\mapsto (X_\mu,D_\mu)$ which connect the noncovariant and natural
realizations of $\hat x_\mu$. One can show that $D_\mu$ is given in terms of $\p_\mu$ according to
\begin{equation}
D_0 = \frac{e^{-\Psi(A)}-1}{ia_0}+\frac{ia_0}{2}\, \square, \quad
D_i = \p_i\, \frac{e^{-\Psi(A)}}{\varphi (A)},  \label{40}
\end{equation}
where $\square$ is the deformed Laplace operator
\begin{equation}\label{58}
\square = \vartriangle \frac{e^{-\Psi(A)}}{\varphi^2(A)}-\left(\frac{2}{ia_0}\right)^2 \sinh^2 \left(\frac{1}{2}\Psi(A)\right), \quad \vartriangle =
\sum_{i=1}^{n-1} \p_i^2.
\end{equation}
The deformed Laplace operator satisfies the commutation relation $[\square, \hat x_\mu]=2D_\mu$. Moreover, the transformation of $X_\mu$ is given by
\begin{align}
X_0 &= \left[x_0 \psi(A)+ia_0 \big(\sum_{k=1}^{n-1} x_k \p_k\big) \gamma(A)\right] \frac{1}{1+\frac{a_0^2}{2}\, \square},  \label{45} \\
X_i &= x_i \varphi(A) e^{\Psi(A)}+ia_0 \left[x_0 \psi(A)+ia_0 \big(\sum_{k=1}^{n-1} x_k \p_k\big) \gamma(A)\right] \frac{1}{1+\frac{a_0^2}{2}\,
\square} \frac{\p_i}{\varphi(A)}.  \label{46}
\end{align}
Substituting Eqs. \eqref{40} and \eqref{45}--\eqref{46} into Eq. \eqref{33} we obtain the noncovariant realization of $M_{\mu\nu}$:
\begin{align}
M_{i0} &= x_i \varphi(A) \left(\frac{1-e^{\Psi(A)}}{ia_0}+\frac{ia_0}{2}\, \square\, e^{\Psi(A)}\right)
-\left[x_0 \psi(A)+ia_0 \Big(\sum_{k=1}^{n-1}x_k \p_k\Big) \gamma (A)\right]\frac{\p_i}{\varphi(A)},  \label{42} \\
M_{ij} &= x_i \p_j - x_j \p_i.  \label{41}
\end{align}
The realizations \eqref{40}--\eqref{58} and \eqref{42}--\eqref{41} generalize the results found in Refs. \cite{Meljanac-2}
and \cite{Borowiec}. For example, if $\psi=1$ or $\psi=1+2A$ we obtain the realizations found in Ref. \cite{Meljanac-2},
and if $\psi=1+rA$, $r\neq 0$, and $\gamma=const.$ we reproduce the realizations found in Ref. \cite{Borowiec} (with $\tau=1$).
The noncovariant realization corresponds to the algebra deformation in example \eqref{99}--\eqref{105}.

\section{Hopf algebra structure of $\kappa$-Poincar\'{e} algebra}

In this section we give a brief description of the Hopf algebra structure of the $\kappa$-Poincar\'{e} algebra $\mathcal{P}_\kappa$.
In the algebra sector the Lorentz generators satisfy the standard relations \eqref{02}, and the commutator $[M_{\mu\nu},p_\lambda]$
in Eq. \eqref{96} is assumed to be deformed by Eqs. \eqref{103}-\eqref{105}. The reason for considering this Hopf algebra
structure is its relation to the differential algebra on the $\kappa$-Minkowski space discussed in section 4. The coproduct and
antipodes of $p_\mu$ and $M_{\mu\nu}$ can be conveniently expressed in terms of the shift operator $Z=e^{\Psi(A)}$ where
$A=a_0 p_0$ (c.f. Eq. \eqref{34}). Since $\Delta Z = Z\otimes Z$ (see Ref. \cite{Meljanac-3}) we find
\begin{equation}\label{63A}
\Delta p_0 = \frac{1}{a_0} \Psi^{-1}\big(\ln (Z\otimes Z)\big)
\end{equation}
where $\ln(Z\tp Z)=\ln(Z)\tp 1 + 1\tp \ln(Z)$. Similarly, one can show that (see Refs. \cite{Meljanac-2} and \cite{Meljanac-4})
\begin{equation}\label{63B}
\Delta p_i = \varphi(a_0 \Delta p_0) \left(\frac{p_i}{\varphi(a_0 p_0)}\otimes 1 + Z\otimes \frac{p_i}{\varphi (a_0 p_0)}\right).
\end{equation}
Furthermore, one finds that the coproducts of the Lorentz generators are given by
\begin{align}
\Delta M_{i0} &= M_{i0}\otimes 1 + Z\otimes M_{i0}-a_0 \sum_{j=1}^{n-1} \frac{p_j}{\varphi (a_0 p_0)}\otimes M_{ij},  \label{65} \\
\Delta M_{ij} &= M_{ij}\otimes 1 + 1\otimes M_{ij}. \label{64}
\end{align}
The counits for all the generators are undeformed. From the definition of antipode \cite{Majid-2} and using Eqs. \eqref{63A}--\eqref{64} we find
\begin{align}
S(p_0) &= \frac{1}{a_0} \Psi^{-1}\big(\ln(Z^{-1})\big), \\
S(p_i) &= -p_i \frac{\varphi(S(a_0 p_0))}{\varphi (a_0 p_0)} Z^{-1},\\
S(M_{i0}) &= -Z^{-1} M_{i0}-a_0 Z^{-1} \sum_{j=1}^{n-1} \frac{p_j}{\varphi (a_0 p_0)} M_{ij},  \label{84}\\
S(M_{ij}) &= -M_{ij}.  \label{70}
\end{align}
The antipode of the shift operator is given by $S(Z)=Z^{-1}$. The coalgebra structure as well
as the antipodes are deformed in all realizations, and particularly in the natural and noncovariant realizations considered here. In the
special case when $\varphi = \psi =1$ we obtain
\begin{alignat}{2}
\Delta p_0 &= p_0 \otimes 1 + 1\otimes p_0, \quad &  \Delta M_{i0} &= M_{i0}\otimes 1 + Z\otimes M_{i0}-a_0 \sum_{j=1}^{n-1} p_j \otimes M_{ij},\\
\Delta p_i &= p_i \otimes 1 + Z\otimes p_i, \quad & \Delta M_{ij} &= M_{ij}\otimes 1 + 1\otimes M_{ij}.
\end{alignat}
Similarly, the antipodes yield
\begin{alignat}{2}
S(p_0) &= -p_0, \quad & S(M_{i0}) &= -Z^{-1} M_{i0}-a_0 Z^{-1} \sum_{j=1}^{n-1} p_j M_{ij}, \\
S(p_i) &= -Z^{-1}p_i, \quad & S(M_{ij}) &= -M_{ij}.
\end{alignat}

Relations \eqref{63A}--\eqref{70} describe the Hopf algebra structure of $\mathcal{P}_\kappa$ in different bases corresponding to different
choices of $\varphi$ and $\psi$. For example, the choice $\varphi=\psi=1$ described above corresponds to the bicrossproduct basis \cite{Majid},
\cite{Sitarz}, while $\varphi=e^{-A}$ and $\psi=1$ corresponds to the left ordering \cite{Dimitrijevic}, \cite{Meljanac-2}. Similarly,
$\varphi=A/(e^A-1)$ and $\psi=1$ corresponds to the Weyl symmetric ordering \cite{Dimitrijevic}, \cite{Meljanac-2}, \cite{Meljanac-3}.
Furthermore, if $\varphi=\psi=1-A$ (resp. $\varphi=1$, $\psi=1+A$) we obtain a basis that corresponds to the left (resp. right) covariant
realization in Refs. \cite{Bu-Kim}, \cite{Meljanac-3} and \cite{Borowiec}. The coproduct and antipode for the generators $P_\mu=-iD_\mu$ in the
natural realization are given in Refs. \cite{Meljanac-3} and \cite{Borowiec-3}. The Hopf algebra structure of $\mathcal{P}_\kappa$ in the
natural realization \eqref{27} is related to the classical basis of $\mathcal{P}_\kappa$ \cite{Kowalski-Glikman-1}, \cite{Borowiec-3}. We note
that the coproducts for $P_\mu$, $N_i$ and $M_i$ used in Ref. \cite{Sitarz} correspond to the coproducts for $p_\mu$ and $M_{\mu\nu}$ when
$\varphi=\psi=1$ in Eqs. \eqref{63A}--\eqref{64}.

\section{Differential forms on $\kappa$-Minkowski space}

Differential calculus on the $\kappa$-deformed Euclidean and Minkowski spaces were considered by several authors in Refs. \cite{Sitarz},
\cite{Dimitrijevic}--\cite{Meljanac-1}. In Ref. \cite{Sitarz} Sitarz has shown that there is no four--dimensional
bicovariant differential caluculus on the $\kappa$-Minkowski space $\Mk$ which is Lorentz covariant. If both conditions are sastisfied
this leads to a contradiction with the mixed Jacobi identity for NC coordinates and one-forms. In order to avoid the problem Sitarz
has constructed a differential calculus in which the space of one-forms is five--dimensional.
However, in an $n$-dimensional spacetime one should expect exactly $n$ one-forms generated by the action of exterior
derivative on the coordinates. In this work we take a different approach based on realizations introduced in section 2.
We show that one can define $n$-dimensional differential algebra on $\Mk$ which is consistent with an action of the
Lorentz algebra in the sense that all graded Jacobi identities involving the NC coordinates, Lorentz generators and one-forms are satisfied.

Let $\hat x_\mu$ be the coordinates on $\Mk$ satisfying relations \eqref{01}, and suppose $\hat x_\mu$ are represented in a $\phi$-realization
\eqref{04}. We introduce deformed exterior derivative $\hat d$ and one-forms $\xi_\mu$ by
\begin{equation}\label{85}
\hat d = \sum_{\a, \beta=0}^{n-1} dx^\a\, \p_\beta k_{\a\beta}(\p), \quad \xi_\mu = \sum_{\a=0}^{n-1} dx^\a\, h_{\a\mu}(\p)
\end{equation}
where $dx^\a = \sum_\beta dx_\beta\, \eta_{\beta \a}$ and $k_{\mu\nu}$, $h_{\mu\nu}$ are formal power series in $a_\mu$ with coefficients in the
ring of differential operators $\p_\mu$. Differential forms $dx_\mu$ satisfy the commutation relations
\begin{equation}
[dx_\mu,x_\nu]=[dx_\mu,\p_\nu]=0 \quad \text{and}\quad \{dx_\mu,dx_\nu\}=0.
\end{equation}
The algebra generated by $x_\mu$, $\p_\mu$ and $dx_\mu$ is a Lie superalgebra graded by the degree of $dx_\mu$
($\text{deg}(x_\mu)=\text{deg}(\p_\mu)=0$ and $\text{deg}(dx_\mu)=1$). The matrix $[h_{\mu\nu}]$ is assumed
to be regular. Furthermore, we assume that $k_{\mu\nu}\to \delta_{\mu\nu}$ and $h_{\mu\nu}\to \eta_{\mu\nu}$ as $a\to 0$, hence in the classical
limit we have $\hat d \to d=\sum_{\a}dx^\a \p_\a$ and $\xi_\mu\to dx_\mu$ as $a\to 0$. Let us define an action of
the exterior derivative on monomials $f(\hat x)$ by $\hat d\cdot \hat f =[\hat d,\hat f]$. We note that in the classical limit we have
$d\cdot x_\mu = [d,x_\mu]=dx_\mu$. Hence, in the noncommutative case we require that $\xi_\mu$ and $\hat d$ are related by
\begin{equation}\label{49}
\xi_\mu = [\hat d,\hat x_\mu].
\end{equation}
Using realizations \eqref{85} and the fundamental relation \eqref{49} we want to construct a differential calculus on $\Mk$
that satisfies the following properties:
\begin{enumerate}[(i)]
\item $\hat d^2=0$,
\item one-forms anti-commute, $\{\xi_\mu,\xi_\nu\}=0$ where $\{\xi_\mu,\xi_\nu\} = \xi_\mu \xi_\nu + \xi_\nu \xi_\mu$,
\item $\hat d$ satisfies the undeformed Leibniz rule
\begin{equation}\label{48}
\hat d\cdot (\hat f\hat g)=(\hat d\cdot \hat f)\hat g + \hat f (\hat d\cdot \hat g)
\end{equation}
where $\hat f$ and $\hat g$ are monomials in $\hat x_\mu$,
\item the commutator $[\xi_\mu,\hat x_\nu]$ is closed in the vector space spanned by one-forms alone,
\begin{equation}\label{47}
[\xi_\mu, \hat x_\nu] = \sum_{\lambda=0}^{n-1} iK^\lambda_{\mu\nu}\, \xi_\lambda, \quad K^\lambda_{\mu\nu}\in \R.
\end{equation}
\end{enumerate}
We note that the commutator $[\xi_\mu,\hat x_\nu]=\xi_\mu \hat x_\nu - \hat x_\nu \xi_\mu$ depends on the realizations of $\hat x_\mu$
and $\xi_\mu$ and need not be closed in $\xi_\mu$.

A generalization of the above construction to higher-order forms was presented in detail in Ref. \cite{Meljanac-1}. Here we only state that
a $k$-form is a finite linear combination of monomials in $\hat x_0,\hat x_1,\ldots ,\hat x_{n-1}$ and $\xi_0,\xi_1,\ldots ,\xi_{n-1}$
such that there are precisely $k$ one-forms in each monomial. One can extend $\hat d$ to a linear map $\hat d\colon
\hat \Omega^k \to \hat \Omega^{k+1}$ where $\hat \Omega^k$ is the space of $k$-forms. In general a $k$-form cannot be written such
that all $\xi_\mu$'s are placed to the far right unless Eq. \eqref{47} holds. If this is true, than any $k$-form is a linear combination of
forms of the type $\hat f_p(\hat x)\xi_{\mu_1}\xi_{\mu_2}\ldots \xi_{\mu_p}$, $0\leq p\leq k$. Furthermore, in this case one can define
an extended star-product of (classical) differential forms \cite{Meljanac-1}.

Relation \eqref{49} is equivalent to a system of PDE's relating the functions $k_{\mu\nu}$, $h_{\mu\nu}$ and $\phi_{\mu\nu}$. Solutions
of such a system in different realizations were discussed in Ref. \cite{Meljanac-1}. Without any further requirements on
$k_{\mu\nu}$ and $h_{\mu\nu}$ the exterior derivative and one-forms satisfy the properties (i)-(iii). We note that consistency
of Eqs. \eqref{01} and \eqref{49} requires that $\xi_\mu$ and $\hat x_\mu$ satisfy the compatiblity condition
\begin{equation}\label{51}
[\xi_\mu,\hat x_\nu]-[\xi_\nu,\hat x_\mu]=i(a_\mu \xi_\nu-a_\nu \xi_\mu).
\end{equation}
This condition places certain restrictions on the realizations of $\xi_\mu$. For example,
$h_{\mu\nu}=\delta_{\mu\nu}$ is not an admissible realization since in this case $[\xi_\mu,\hat x_\nu]=0$ for all $\mu,\nu=0,1,\ldots ,n-1$
contradicting Eq. \eqref{51}.

Let us consider condition (iv). In general, $K_{\mu\nu}^\lambda$ is a formal power series in $\p_\mu$ and it depends on the realizations
of $\hat x_\mu$ and $\xi_\mu$. Using Eq. \eqref{51} one can decompose $K^\lambda_{\mu\nu}$ into symmetric and antisymmetric parts
\begin{equation}
K^\lambda_{\mu\nu}=A^\lambda_{\mu\nu}+S^\lambda_{\mu\nu},
\end{equation}
where $A^\lambda_{\mu\nu}=\frac{1}{2}(a_\mu \delta_{\nu\lambda}-a_\nu \delta_{\mu\lambda})$ and
\begin{equation}
S^\lambda_{\mu\nu}=-\frac{i}{2}\sum_{\a,\beta=0}^{n-1}h^{-1}_{\lambda\a}\left(\frac{\p h_{\a\mu}}{\p \p_\beta}\phi_{\beta\nu}+ \frac{\p
h_{\a\nu}}{\p \p_\beta}\phi_{\beta\mu}\right).
\end{equation}
Here $h^{-1}_{\mu\nu}$ denotes the $(\mu,\nu)$ element of the inverse matrix $[h_{\mu\nu}]^{-1}$. Thus, in order to satisfy condition (iv)
we need to find $h_{\mu\nu}$ such that the symmetric part $S^\lambda_{\mu\nu}$ is constant.
Solving the above problem in full generality is fairly complicated. However, by way of a concrete example we show that such solutions exist.
For a given noncovariant realization of $\hat x_\mu$ we will construct $\hat d$ such that the one-forms given by Eq. \eqref{49}
have the desired properties.

Assume the following Ansatz for $\hat d$:
\begin{equation}
\hat d = -dx_0 \p_0 K_1(A)+\Big(\sum_{k=1}^{n-1} dx_k \p_k\Big) K_2(A), \quad A=-ia_0\p_0.
\end{equation}
Using the realization \eqref{29}--\eqref{30} for $\hat x_\mu$ we find
\begin{align}
\xi_0 &= [\hat d,\hat x_0] = dx_0 (AK_1^\prime +K_1)\psi +ia_0\Big(\sum_{k=1}^{n-1} dx_k \p_k\Big) (\psi  K_2^\prime +\gamma K_2), \label{52} \\
\xi_i &= [\hat d,\hat x_i]=dx_i K_2 \varphi,  \label{53}
\end{align}
where $K_i^\prime = \frac{dK_i}{dA}$. We want to find $K_1$ and $K_2$ such that $\xi_0=dx_0 Z^{-s}$ and $\xi_i = dx_i Z^{-t}$ for some $s,t\in \R$,
where the shift operator $Z$ is given by Eq. \eqref{34}. From Eqs. \eqref{52} and \eqref{53} we obtain a system of differential equations
\begin{equation}
(AK_1^\prime + K_1) \psi = Z^{-s}, \quad \psi K_2^\prime +\gamma K_2 = 0, \quad \varphi K_2 = Z^{-t}. \label{55}
\end{equation}
Since $\gamma = \psi\varphi^\prime /\varphi+1$, the last two equations are compatible if and only if $t=1$. Hence,
\begin{equation}
K_2(A) = \frac{Z^{-1}}{\varphi (A)}.
\end{equation}
Solving the differential equation for $K_1$ and taking into account the initial condition\\ $\lim_{a_0\to 0} K_1(A)=1$ yields
\begin{equation}
K_1(A)=\frac{1-Z^{-s}}{sA}, \quad s\neq 0.
\end{equation}
In the limit $s\to 0$ the solution is given by
\begin{equation}
K_1(A)=\frac{1}{A}\int_0^A \frac{dt}{\psi(t)}.
\end{equation}
Thus, we obtain a one-parameter family of exterior derivatives
\begin{equation}\label{66}
\hat d = -dx_0 \p_0\, \frac{1-Z^{-s}}{sA}+\Big(\sum_{k=1}^{n-1} dx_k \p_k\Big) \frac{Z^{-1}}{\varphi (A)},
\end{equation}
and corresponding one-forms
\begin{equation}
\xi_0 = dx_0 Z^{-s}, \quad \xi_i = dx_i Z^{-1}.
\end{equation}
Since the shift operator satisfies $[Z^\a, \hat x_\mu]=\a\, ia_\mu Z^\a$, $\a\in \R$, it follows that the commutators $[\xi_\mu,\hat x_\nu]$ are
closed:
\begin{alignat}{2}
[\xi_0, \hat x_0] &= -sia_0 \xi_0, \quad & [\xi_i,\hat x_0] &= -ia_0 \xi_i,   \label{56} \\
[\xi_0,\hat x_j] &=0, \quad & [\xi_i,\hat x_j] &=0.  \label{57}
\end{alignat}
We point out that the algebra generated by $\hat x_\mu$ and $\xi_\mu$ is closed for all noncovariant realizations \eqref{29}--\eqref{30}
and that all graded Jacobi identities for this algebra hold.

Let us now consider the commutation relations for $M_{\mu\nu}$ and $\xi_\lambda$. Using the natural realization \eqref{27} one can
express the Lorentz generators as
\begin{equation}
M_{\mu\nu}=(\hat x_\mu D_\nu - \hat x_\nu D_\mu)Z
\end{equation}
which yields
\begin{equation}
[M_{\mu\nu},\xi_\lambda]=[\hat x_\mu,\xi_\lambda] D_\nu Z -[\hat x_\nu,\xi_\lambda]D_\mu Z.
\end{equation}
Thus, one may use the commutation relations \eqref{56}--\eqref{57} to find
\begin{alignat}{2}
[M_{i0},\xi_0] &= -sia_0 \xi_0\, \frac{\p_i}{\varphi (A)}, \quad & [M_{ij},\xi_0] &= 0, \\
[M_{i0}, \xi_k] &= -ia_0 \xi_k\, \frac{\p_i}{\varphi (A)}, \quad & [M_{ij},\xi_k] &=0.
\end{alignat}
The algebra generated by $\hat x_\mu$, $\xi_\mu$ and $M_{\mu\nu}$ is not closed because the commutator $[M_{i0},\xi_\mu]$
is given in terms of an infinite power series in $\p_\mu$. However, since $\hat x_\mu$, $M_{\mu\nu}$
and $\xi_\mu$ belong to an associative algebra generated by $x_\mu$, $\p_\mu$ and $dx_\mu$ all graded Jacobi identities hold.
Thus, one can define an action of $M_{\mu\nu}$ on the differential algebra $\Dk$ defined by
relations \eqref{28}, \eqref{56} and \eqref{57} as follows. First define the action of $\hat x_\mu$
and $\xi_\mu$ on $\Dk$ simply by $\hat x_\mu \rhd f(\hat x,\xi)=\hat x_\mu f(\hat x,\xi)$ and $\xi_\mu \rhd f(\hat x,\xi)=
\xi_\mu f(\hat x,\xi)$ for all $f(\hat x,\xi)\in \Dk$. Furthermore, define $M_{\mu\nu}\rhd 1 =0$. Since the commutator
$[M_{\mu\nu},\xi_\lambda]$ depends on $\p_\mu$ we also need to set $\p_\mu \rhd 1 = 0$. Now we define the action of $M_{\mu\nu}$ on
$\Dk$ by
\begin{equation}\label{59}
M_{\mu\nu}\rhd f(\hat x,\xi) = \big(M_{\mu\nu} f(\hat x,\xi)\big)\rhd 1.
\end{equation}
The action \eqref{59} is completely specified by the action of $\hat x_\mu$, $\xi_\mu$, $M_{\mu\nu}$ and $\p_\mu$, and the
commutation relations between $M_{\mu\nu}$, $\hat x_\mu$ and $\xi_\mu$. Also, due to the Jacobi identities
the action \eqref{59} is compatible with the commutation relations \eqref{56}--\eqref{57}. Since $M_{\mu\nu}\rhd 1=0$,
Eq. \eqref{59} can be written in equivalent form
\begin{equation}\label{86}
M_{\mu\nu}\rhd f(\hat x,\xi) = [M_{\mu\nu}, f(\hat x,\xi)]\rhd 1.
\end{equation}
When the action is restricted to monomials in $\hat x_\mu$, due to commutation relations \eqref{03} one obtains a
polynomial in $\hat x_\mu$, witten symbolically
\begin{equation}\label{82}
M_{\mu\nu}\rhd f(\hat x) = g(\hat x),
\end{equation}
and the result is clearly independent of realizations. For example, the action of $M_{\mu\nu}$ on $\hat x_\mu$ yields
\begin{alignat}{2}
M_{i0} \rhd \hat x_0 &= -\hat x_i, \quad & M_{ij} \rhd \hat x_0 &= 0, \label{61} \\
M_{i0} \rhd \hat x_k &= -\delta_{ik}\hat x_0, \quad & M_{ij} \rhd \hat x_k &= \delta_{jk} \hat x_i- \delta_{ik}\hat x_j, \label{62}
\end{alignat}
The above result agrees with the action of the Lorentz generators on $\kappa$-Minkowski space obtained by Sitarz \cite{Sitarz}.
Furthermore,
\begin{equation}\label{83}
M_{\mu\nu} \rhd f(\xi) = 0
\end{equation}
for any monomial $f(\xi)$ in $\xi_\mu$. Since a basis of $\Dk$ consists of the monomials
\begin{equation}
\hat x_0^{k_0}\, \hat x_1^{k_1}\ldots \hat x_{n-1}^{k_{n-1}}\, \xi_0^{l_0}\, \xi_1^{l_1}\ldots \xi_{n-1}^{l_{n-1}}, \quad
k_i \geq 0, \quad l_i=0,1,
\end{equation}
it suffices to calculate the action of $M_{\mu\nu}$ on the product of monomials $f(\hat x)g(\xi)$.
Using Eqs. \eqref{82} and \eqref{83} one can show that
\begin{equation}\label{88}
M_{\mu\nu}\rhd \big(f(\hat x)g(\xi)\big) = \big(M_{\mu\nu} \rhd f(\hat x)\big) g(\xi)
\end{equation}
which is again independent of realization. Thus, the action of $M_{\mu\nu}$ on the entire differential algebra $\Dk$ is realization independent.

The action \eqref{88} can be expressed in terms of the quantum adjoint action
\begin{equation}
\text{ad}(M_{\mu\nu})(f(\hat x)) = \sum M_{{\mu\nu}_{(1)}} f(\hat x) S\big(M_{{\mu\nu}_{(2)}}\big)
\end{equation}
where we use the Sweedler notation for the coproduct $\Delta M_{\mu\nu}=\sum M_{{\mu\nu}_{(1)}}\otimes M_{{\mu\nu}_{(2)}}$. Since the rotation
generators are undeformed (cf. Eqs. \eqref{64} and \eqref{70}), we have
\begin{equation}
\text{ad}(M_{ij})(f(\hat x))=[M_{ij},f(\hat x)].
\end{equation}
The coproduct and antipode for boosts given by Eqs. \eqref{65} and \eqref{84} yield
\begin{equation}\label{87}
\begin{split}
\text{ad}(M_{i0})(f(\hat x)) &= M_{i0}f(\hat x)-Z f(\hat x) Z^{-1} M_{i0} \\
&+ia_0 Zf(\hat x)Z^{-1} \sum_{j=1}^{n-1} \frac{\p_j}{\varphi (A)} M_{ij} - ia_0 \sum_{j=1}^{n-1} \frac{\p_j}{\varphi (A)} f(\hat x)M_{ij}.
\end{split}
\end{equation}
where we have used $p_\mu = -i\p_\mu$.
If $f(\hat x)$ is a monomial of degree $m$, then Eq. \eqref{92} implies $Zf(\hat x)Z^{-1}=f(\hat x+ia)=f(\hat x)+r_{m-1}(\hat x)$
where $r_{m-1}(\hat x)$ is a monomial of degree $m-1$. Subtituting this into Eq. \eqref{87} we obtain
\begin{equation}
\begin{split}
\text{ad}(M_{i0})(f(\hat x)) &= [M_{i0},f(\hat x)]-r_{m-1}(\hat x)M_{i0} \\
&+ia_0 \big(f(\hat x)+r_{m-1}(\hat x)\big) \sum_{j=1}^{n-1} \frac{\p_j}{\varphi (A)}M_{ij} -ia_0 \sum_{j=1}^{n-1}
\frac{\p_j}{\varphi (A)} f(\hat x)M_{ij}.
\end{split}
\end{equation}
Since $\p_\mu \rhd 1=0$ and $M_{\mu\nu}\rhd 1=0$, it follows that
\begin{equation}
M_{\mu\nu}\rhd f(\hat x) = [M_{\mu\nu},f(\hat x)] \rhd 1 = \text{ad}(M_{\mu\nu}) \big(f(\hat x)\big) \rhd 1,
\end{equation}
and consequently
\begin{equation}
M_{\mu\nu}\rhd f(\hat x)g(\xi) = \Big(\text{ad}(M_{\mu\nu})(f(\hat x))\rhd 1\Big) g(\xi).
\end{equation}

A few comments about the action \eqref{59} are in order. Although the action is independent of the realizations introduced
in section 2, this is generally not true if the commutator $[M_{\mu\nu},\hat x_\lambda]$ given by Eq. \eqref{03} is modified.
The action \eqref{59} is different from the one introduced by Sitarz \cite{Sitarz},
but both actions agree when restricted to the coordinates of the $\kappa$-Minkowski space $\Mk$ (c.f. Eqs. \eqref{61}--\eqref{62}).
We remarked earlier that when the action in \cite{Sitarz} is extended in a covariant way from $\Mk$ to the differential algebra $\Dk$
one obtains a contradiction with the mixed Jacobi identity for $\hat x_\mu$, $\hat x_\nu$ and $\xi_\lambda$. In order to resolve the
contradiction Sitarz introduced an additional one-form $\phi$ which is Lorentz invariant, $M_{\mu\nu}\rhd \phi =0$,
thus making the space of one-forms $(n+1)$-dimensional. In Ref. \cite{Bu-Kim} the same problem was resolved by extending
the $\kappa$-Poincar\'{e} algebra with a dilatation operator and using a different coproduct for $M_{\mu\nu}$.
In this case the commutator $[M_{\mu\nu},\hat x_\lambda]$ is different from the one in Eq. \eqref{03}. The coproduct for
$P_\mu$ in Ref. \cite{Bu-Kim} corresponds to the coproduct for $\p_\mu$ in the left-covariant realization in Ref. \cite{Meljanac-3} and
to the special case of the noncovariant realization \eqref{29}--\eqref{30} with $\varphi = \psi=1-A$.
We note that the requirement $[M_{\mu\nu},\hat d]=0$ is equivalent to the bicovariance requirement in Ref.
\cite{Sitarz}. In our work, however, $[M_{\mu\nu},\hat d]\neq 0$ for any choice of $\varphi$ and $\psi$ in the noncovariant realization.
We also note that this is true even in the classical limit since $[M_{\mu\nu},\hat d]\to dx_\mu \p_\nu - dx_\nu \p_\mu\neq 0$ as $a\to 0$.

We conclude the discussion with the following remarks. In this paper we have generalized the realizations of $\hat x_\mu$ and $M_{\mu\nu}$
introduced originally in Ref. \cite{Meljanac-2}. We have given the Hopf algebra structure of the $\kappa$-Poincar\'{e} algebra
$\mathcal{P}_\kappa$ when the deformation of the algebra sector is given by Eqs. \eqref{103}-\eqref{105}. In addition, we found realizations of
the exterior derivative $\hat d$ and one-forms $\xi_\mu$ such that $[\xi_\mu,\hat x_\nu]$ is closed in the vector space spanned by one-forms
alone. The requirement that both commutators $[M_{\mu\nu},\hat x_\lambda]$ and $[\xi_\mu,\hat x_\nu]$ be closed is not compatible with Lorentz
invariance of $\hat d$ since $[M_{\mu\nu},\hat d]\neq 0$. A different construction of the Lorentz generators satisfying the invariance condition
$[M_{\mu\nu},\hat d]=0$ will be presented elsewhere. Full treatment of this construction as well as its applications to Snyder space
\cite{Battisti}--\cite{Meljanac-5}, scalar field theory, statistics and twist operators \cite{Bu-Kim}, \cite{Borowiec-2}, \cite{Borowiec},
\cite{Govindarajan}, \cite{Govindarajan-2} will be given in future work.

\end{document}